\documentclass[aps,prl,preprint]{revtex4-1}
\usepackage{graphicx}
\usepackage{amsfonts}
\usepackage{textcomp}
\usepackage{verbatim}
\usepackage {ulem}
\newcommand{\cmfs}{Co$_{\mathrm{2}}$Mn$_{\mathrm{0.6}}$Fe$_{\mathrm{0.4}}$Si}

\newcommand{\mum}{$\mu$m}
\newcommand{\mubls}{$\mu$BLS}
\newcommand{\hext}{\muzero \ensuremath{H_{\mathrm{ext}}}}
\newcommand{\heff}{\ensuremath{H_{\mathrm{eff}}}}

\newcommand{\muzero}{\ensuremath{\mu_{\mathrm{0}}}}

\newcommand{\nife}{Ni$_\mathrm{81}$Fe$_\mathrm{19}$}
\begin{document}

\title{Nonlinear emission of spin-wave caustics from an edge mode of a micro-structured Co$_{\mathrm{2}}$Mn$_{\mathrm{0.6}}$Fe$_{\mathrm{0.4}}$Si waveguide}

\author{T.\,Sebastian}

\email{tomseb@physik.uni-kl.de}

\affiliation{Fachbereich Physik and Forschungszentrum OPTIMAS, Technische Universit\"at
Kaiserslautern, 67663 Kaiserslautern, Germany}

\affiliation{Graduate School Materials  Science in Mainz, Gottlieb-Daimler-Stra\ss e 47,
67663 Kaiserslautern, Germany}

\author{P.\,Pirro}

\affiliation{Fachbereich Physik and Forschungszentrum OPTIMAS, Technische Universit\"at
Kaiserslautern, 67663 Kaiserslautern, Germany}

\author{T.\,Br\"{a}cher}

\affiliation{Fachbereich Physik and Forschungszentrum OPTIMAS, Technische Universit\"at
Kaiserslautern, 67663 Kaiserslautern, Germany}

\affiliation{Graduate School Materials  Science in Mainz, Gottlieb-Daimler-Stra\ss e 47,
67663 Kaiserslautern, Germany}

\author{T.\,Kubota}

\affiliation{WPI Advanced Institute for Materials Research, Tohoku University, Katahira 2-1-1, Aoba-ku, Sendai 980-8577, Japan}

\author{A.A.\,Serga}

\affiliation{Fachbereich Physik and Forschungszentrum OPTIMAS, Technische Universit\"at
Kaiserslautern, 67663 Kaiserslautern, Germany}

\author{H.\,Naganuma}

\affiliation{Department of Applied Physics, Graduate School of Engineering, Tohoku University, Aoba-yama 6-6-05, Sendai 980-8579, Japan}

\author{M.\,Oogane}
\affiliation{Department of Applied Physics, Graduate School of Engineering, Tohoku University, Aoba-yama 6-6-05, Sendai 980-8579, Japan}

\author{Y.\,Ando}

\affiliation{Department of Applied Physics, Graduate School of Engineering, Tohoku University, Aoba-yama 6-6-05, Sendai 980-8579, Japan}

\author{B.\,Hillebrands}

\affiliation{Fachbereich Physik and Forschungszentrum OPTIMAS, Technische Universit\"at
Kaiserslautern, 67663 Kaiserslautern, Germany}
\date{\today}

\begin{abstract}
Magnetic Heusler materials with very low Gilbert damping are expected to show novel magnonic transport phenomena. We report nonlinear generation of higher harmonics leading to the emission of caustic spin-wave beams in a low-damping, micro-structured \cmfs{} Heusler waveguide. The source for the higher harmonic generation is a localized edge mode formed by the strongly inhomogeneous field distribution at the edges of the spin-wave waveguide. The radiation characteristics of the propagating caustic waves observed at twice and three times the excitation frequency are described by an analytical calculation based on the anisotropic dispersion of spin waves in a magnetic thin film.

\end{abstract}

\maketitle

In the last years nonlinear spin dynamics in magnetic microstructures made of metallic ferromagnetic thin films or layer stacks have gained large interest.\cite{STNO_higher_harmonics, 2nd_harmonic_waveguide, harmonics_dot, 4magnon1, 4magnon2} The intrinsically nonlinear Landau-Lifshitz and Gilbert equation (LLG), that governs the spin dynamics, gives rise to a variety of nonlinear effects.\cite{llg, nonlinear}

Among the metallic ferromagnets, the class of Cobalt-based Heusler materials is very promising for future \textit{magnon spintronics} devices and the observation of new phenomena of magnonic transport. The reasons for the interest in these materials are the small magnetic Gilbert damping, the high spin-polarization, and the high Curie temperature.\cite{permalloy, trudel}

As shown recently, the full Heusler compound \cmfs{} (CMFS) is a very suitable material to be used as a micro-structured spin-wave waveguide due to the increased decay length which was observed for wave propagation in the linear regime.\cite{sebastian} The reason for this observation is the low Gilbert damping of $\alpha=3 \times 10^{-3} $ of CMFS compared to \nife{} with $\alpha=8\times10^{-3}$, which is the material commonly used in related studies.\cite{trudel} The decreased magnetic losses not only lead to an increase of the decay length but also to large precession angles of the magnetic moments and, thus, to the occurrence of nonlinear effects. Regarding future applications, the investigation and the thorough understanding of phenomena related to the spin-wave propagation in the nonlinear regime in Heusler compounds is crucial.

In this Letter, we report nonlinear higher harmonic generation from a localized edge mode \cite{edge_mode_1, edge_mode_2} causing the emission of caustic spin-wave beams \cite{kaustic_Py, kaustic_YIG} in a micro-structured Heusler waveguide. Spin-wave caustics are characterized by the small transversal aperture of a beam, which practically does not increase during propagation, and the well-defined direction of propagation.

\begin{figure}
	\centering
		\includegraphics{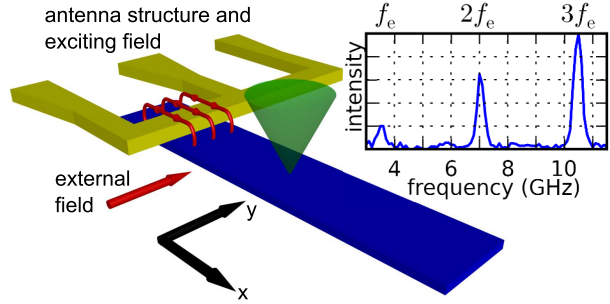}
	\caption{(Color online) Sketch of the sample design. The shortened end of a coplanar waveguide is used as an antenna structure to excite spin dynamics in a 5\,\mum{} wide CMFS waveguide with a thickness of 30\,nm. The waveguide is positioned in the x-y-plane with its long axis pointing along the x-direction. The external magnetic field is applied transversely to the waveguide in y-direction. The figure includes a \mubls{} spectrum taken at a distance of 4.5\,\mum{} from the antenna in the center of the waveguide (see laser beam in the sketch) for an excitation frequency of $f_e=3.5$\,GHz, a microwave power of 20\,mW, and an external field $\hext{}=48$\,mT.}
	\label{sketch_final}
\end{figure}

The investigated sample is a 5\,\mum{} wide spin-wave waveguide structured from a 30\,nm thick film of the Heusler compound CMFS. Details about the fabrication and the material properties can be found in Refs.\,\onlinecite{cmfs_1, cmfs_2}. The microfabrication of the waveguide was performed using electron-beam lithography and ion-milling. For the excitation of spin dynamics in the waveguide the shortened end of a coplanar waveguide made of copper was placed on top of it. The Oersted field created by a microwave current in this antenna structure can be used to excite spin dynamics in the Gigahertz range. The antenna has a thickness of 400\,nm and a width of $\Delta x = 1$\,\mum{}.

All observations have been carried out using Brillouin light scattering microscopy (\mubls{}).\cite{uBLS} \mubls{} is a powerful tool to investigate spin dynamics in microstructures with a spatial resolution of about 250\,nm and a frequency resolution of up to 50\,MHz.

In the following description, the waveguide is positioned in the x-y-plane with the long axis pointing in x-direction. The origin of the coordinate system is given by the position of the antenna between $x=-1$\,\mum{} and $x=0$\,\mum{}. An external magnetic field of $\hext=48$\,mT was applied transversely to the waveguide in y-direction resulting in Damon-Eshbach geometry \cite{DE} for spin waves propagating along the waveguide. A sketch of the sample layout is shown in Fig.\,\ref{sketch_final}.

In addition, Fig.\,\ref{sketch_final} includes a spectrum taken by \mubls{} for an excitation frequency of $f_\mathrm{e}=3.5$\,GHz and a microwave power of 20\,mW at a distance of 4.5\,\mum{} from the antenna in the center of the waveguide. The spectrum shows not only a peak at $f_\mathrm{e}=3.5$\,GHz but also at $2f_\mathrm{e}=7.0$\,GHz and $3f_\mathrm{e}=10.5$\,GHz. Furthermore, the intensity of the directly excited spin wave at 3.5\,GHz is lower than for the higher harmonics at the point of observation. As we will see, the higher harmonics are excited resonantly by nonlinear magnon-magnon interactions and the intensity distribution is a consequence of the different propagation characteristics of the observed spin-wave modes.

\begin{figure}
	\centering
\includegraphics{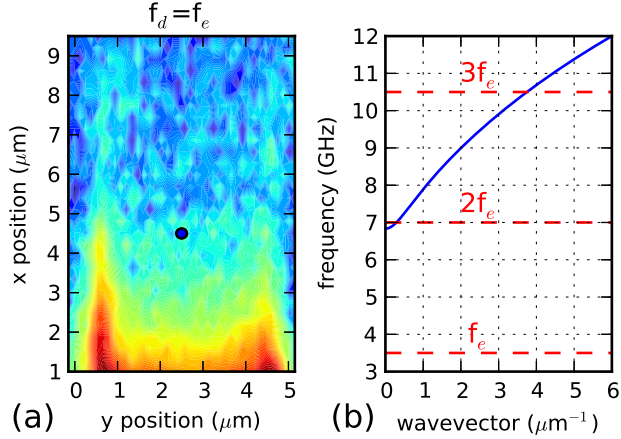}
	\caption{(Color online) (a) \mubls{} intensity distribution for the detection frequency $f_\mathrm{d}$ being equal to the excitation frequency $f_\mathrm{e}=3.5$\,GHz. The observed intensity of the edge mode is maximal near the edges of the CMFS waveguide. Please note that the exciting antenna is positioned between $x = -1$\,\mum{} and $x=0$\,\mum{}. The blue dots in the graph indicates the position of the measurement presented in Fig.\,\ref{sketch_final}. (b) Calculated dispersion relation for the center of the CMFS waveguide according to Ref.\,\onlinecite{dispersion} as well as excitation frequency and higher harmonics (dashed lines).}
	\label{dispersion_intensity}
\end{figure}

A two-dimensional intensity distribution for the detection frequency $f_\mathrm{d}=f_\mathrm{e}=3.5$\,GHz as well as the calculated dispersion relation \cite{dispersion} for the center of the CMFS waveguide are shown in Fig.\,\ref{dispersion_intensity}. Figure\,\ref{dispersion_intensity}(a) reveals a strong localization of the intensity at the edges of the waveguide. The non-vanishing intensity close to the antenna and between the edges of the waveguide ($y=1-4$\,\mum{}) can be attributed to nonresonant, forced excitation by the Oersted field created by the microwave current. Figure\,\ref{dispersion_intensity}(b) shows the spin-wave dispersion for the CMFS waveguide calculated according to Ref.\,\onlinecite{dispersion} assuming a homogeneous magnetization oriented in y-direction by the external field. The material parameters used in all our calculations were determined experimentally on the unstructured film via ferromagnetic resonance ($M_\mathrm{S}=1003$\,kA/m and $H_\mathrm{ani} = 1$\,kA/m) and via \mubls{} in the micro-structured waveguide ($A_\mathrm{ex}=13$\,pJ/m) following a method described in \cite{trudel}. The effective field of $\muzero\heff=46$\,mT used in the calculations was obtained by micromagnetic simulations. As can be seen, the lower cut-off frequency $f = 6.9$\,GHz is well above the excitation frequency of $f_\mathrm{e}=3.5$\,GHz (see dashed line in Fig.\,\ref{dispersion_intensity}(b)). In the center of the waveguide, the assumption of a homogeneous magnetization is a very good approximation. Close to the edges, demagnetizing effects are responsible for a strongly decreased effective field and an inhomogeneous magnetization configuration. The inhomogeneity of the magnetization does not allow for a quantitative modeling of spin dynamics close to edges.\cite{kostylev} However, from previous work it is known that this field and magnetization configuration allow for the existence of localized spin waves - commonly referred to as edge modes - energetically far below the spin-wave dispersion for propagating modes in the center.\cite{edge_mode_1, edge_mode_2} Therefore, we conclude that the spin-wave mode at $f_\mathrm{d}=f_\mathrm{e}=3.5$\,GHz is excited resonantly by the microwave field.

\begin{figure}
	\centering
		\includegraphics{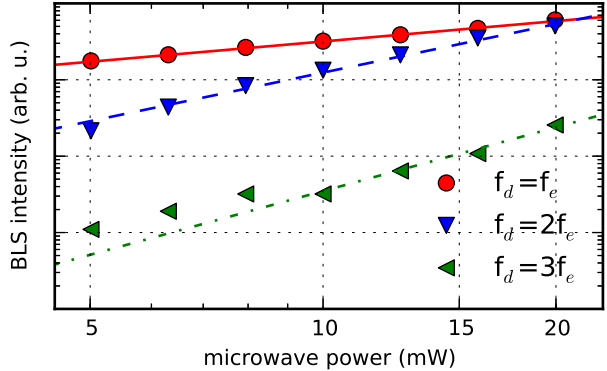}
		\caption{(Color online) Power dependence of the BLS intensity for the detection frequencies $f_\mathrm{d}=3,5$\,GHz, 7\,GHz, and 10.5\,GHz for the fixed excitation frequency $f_\mathrm{e}=3.5$\,GHz. Please note the log-log presentation of the data. The lines in the graph correspond to fits according to Eq.\,\ref{leistung}. A least square fit of the data yields $s_\mathrm{1f}=0.9\pm0.1$, $s_\mathrm{2f}=2.1\pm0.1$, and $s_\mathrm{3f}=2.8\pm0.3$.}
	\label{leistungsabhaengigkeit}
\end{figure}

As can be seen from the spectrum in Fig.\,\ref{sketch_final}, the applied microwave power of 20\,mW is sufficiently high to observe the nonlinear generation of higher harmonics of the excitation frequency $f_\mathrm{e}$ in the \mubls{} spectra. These resonant frequency multiplications to $f=2f_\mathrm{e}=7.0$\,GHz and $f=3f_\mathrm{e}=10.5$\,GHz result in the excitation of propagating spin-wave modes in the waveguide energetically above the cut-off frequency of the dispersion shown in Fig.\,\ref{dispersion_intensity}(b).

Figure\,\ref{leistungsabhaengigkeit} shows the dependence of the directly excited mode and the higher harmonics on the applied microwave power. This data has been acquired close to the position of the edge mode and near the antenna at $x=0.7$\,\mum{} and $y=0.8$\,\mum{}. The data is presented on a log-log scale with fits according to

\begin{equation}
	I_n(p)=A_np^{s_n}+b,
	\label{leistung}
\end{equation}

where $I_n$ is the \mubls{} intensity, $A_n$ a coupling parameter, $p$ the applied microwave power, and $b$ the noise-level in our measurement. As expected, these processes do not show a threshold power level, but reliable detection on the background of the noise is not possible for powers below 5\,mW for $f_\mathrm{d}=3f_\mathrm{e}=10.5$\,GHz. The different slopes of the curves for the different spin-wave modes $n$ are caused by the different power-laws specified by the exponent $s_n$. A least square fit of the data yields $s_\mathrm{1f}=0.9\pm0.1$, $s_\mathrm{2f}=2.1\pm0.1$, and $s_\mathrm{3f}=2.8\pm0.3$. These experimental findings close to the integer values 1, 2, and 3 are in accordance with both reported experimental data and theoretical predictions for the nonlinear generation of higher harmonics.\cite{harmonics_dot, nonlinear} 

The observation of the second harmonic can be understood qualitatively by considering the strong demagnetizing fields caused by the out-of-plane component $m_\mathrm{z}(t)$ during the magnetization precession. Due to the demagnetizing fields, the magnetization precession $\mathbf{M}(t)$ around the y-direction (defined by the effective field) follows an elliptical trajectory rather than a circular one. In contrast to the case of a circular precession, the resulting projection of $\mathbf{M}$ on the y-axis is time dependent and oscillating with the frequency $2f_\mathrm{e}$. The resulting dynamic dipolar field $\left|\mathbf{h}_\mathrm{y}(t) \right| \propto m_{x}^2-m_{z}^2$ can be regarded as the source for the frequency doubling. Similar considerations lead to the observation of higher harmonics. A full quantitative derivation of higher harmonic generation and other nonlinear effects based on the expansion of the LLG in terms of the dynamic magnetization can be found in Ref.\,\onlinecite{nonlinear}.

\begin{figure}
	\centering
		\includegraphics{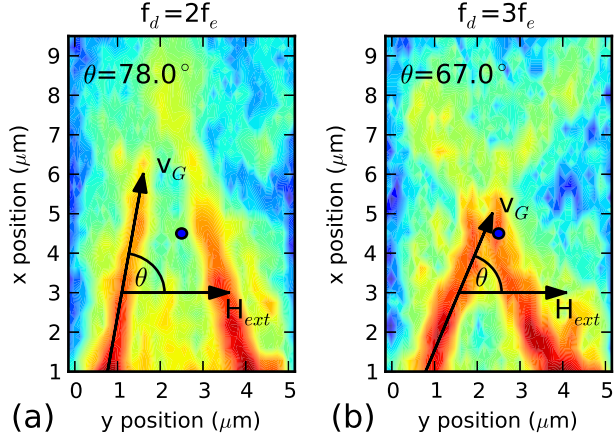}
	\caption{(Color online) \mubls{} intensity distribution for (a) $f_\mathrm{d}=2f_\mathrm{e}=7.0$\,GHz and (b) $f_\mathrm{d}=3f_\mathrm{e}=10.5$\,GHz. Both intensity maps show strongly directed spin-wave beams along the angle $\theta= \angle(\mathbf{H_{ext}},\mathbf{v_G})$. The lines in the maps are guides to the eye to identify the propagation angle $\theta$. The blue dots in the graphs indicate the position of the measurement presented in Fig.\,\ref{sketch_final}.}
	\label{intensity_maps}
\end{figure}

Figures\,\ref{intensity_maps}(a) and (b) show intensity maps for the detection frequencies $f_\mathrm{d}=2f_\mathrm{e}=7.0$\,GHz and $f_\mathrm{d}=3f_\mathrm{e}=10.5$\,GHz, respectively. In both cases the intensity radiated from the position of the edge mode is strongly directed, has a small transversal aperture, and shows nondiffractive behavior. These radiation characteristics recorded for the three spin-wave modes at $f=3.5$, 7.0 and 10.5\,GHz presented in Figs.\,\ref{dispersion_intensity}(a) and \ref{intensity_maps} are responsible for the intensity distribution shown in the spectrum in Fig.\,\ref{sketch_final}. The position of this measurement is indicated in the corresponding intensity maps with a circle. Since the spin wave at $f_\mathrm{e}=3.5$\,GHz is localized at the edges of the waveguide, its intensity is comparably weak in the center. In contrast, the higher harmonics have frequencies above the cut-off frequency of the spin-wave dispersion and can propagate in the center of the waveguide. The propagation angles of these modes support the intensity distribution recorded in our measurement. While an increased intensity can be found already for $f_\mathrm{d}=2f_\mathrm{e}$, for $f_\mathrm{d}=3f_\mathrm{e}$ the two beams starting from both edges of the spin-wave waveguide even intersect at the measurement position resulting in the highest intensity at this point. Because of the well-defined propagation angles of the higher harmonics and the localization of the edge mode, the intensity distribution is strongly depending on the measurement position.

This observation of spin-wave beams with small transversal aperture is reminiscent of the results in Refs.\,\onlinecite{3magnon1, 3magnon2}, where nonlinear three-magnon scattering in yttrium iron garnet is reported. However, in that case, the propagation direction of the nonlinearly generated spin wave is given by momentum conservation in the scattering process. In contrary, in our case, due to the strong localization of the edge mode, the assumption of a well-defined initial wavevector and, thus, a strict momentum conservation is not justified. In particular, it is not possible to find an initial wavevector that allows for the nonlinear generation of the second and third harmonic at the same time still respecting momentum conservation.

In the following, we will describe the observed propagation characteristics using the properties of the anisotropic spin-wave dispersion in a magnetic thin film.\cite{dispersion, kaustic_Py, kaustic_YIG} Because of this anisotropy, the direction of the flow of energy, which is given by the direction of the group velocity $\mathbf{v_{G}}=2\pi \partial f(\mathbf{k})/\partial \mathbf{k}$ of the investigated spin waves, can differ significantly from the direction of its wavevector $\mathbf{k}$. To estimate the relevant range in k-space in our experiment, we have to consider the lateral dimensions of the source for the nonlinear processes. Since the excitation by the oscillating Oersted field is most efficient directly below the antenna, the edge mode has the highest intensity in this region given by the width of the antenna of $\Delta x=1$\,\mum{}. The spread of the edge mode in y-direction can be estimated from the intensity map in Fig.\,\ref{dispersion_intensity}(a) to be smaller than 1\,\mum{}. Because of this localization, the edge mode in our measurement at $f=3.5$\,GHz can be regarded as a source for the nonlinear emission of the higher harmonics with lateral dimensions of approximately $1 \times 1$\,\mum{}$^2$. As a first approximation, the Fourier transformation of this geometry lets us estimate the maximum wavevector that can be excited by the edge mode to be $k_{max}\approx 6.3$\,\mum{}$^{-1}$. As we will see, the direction of the group velocity can be assumed to be constant for most wavevectors that can be excited. This finally leads to the formation of the caustics in our experiment.

\begin{figure}
	\centering
		\includegraphics{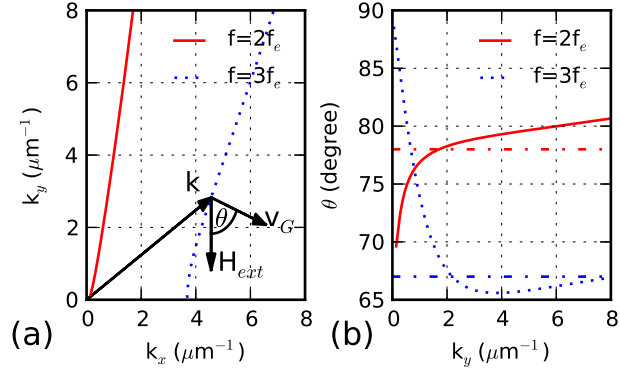}
	\caption{(Color online) Analytical calculations according to Ref.\,\onlinecite{dispersion}. (a) Iso-frequency curves $f_{const}=f(k_x, k_y)$  for $f=2f_\mathrm{e}=7.0$\,GHz and $f=3f_\mathrm{e}=10.5$\,GHz. Based on these calculations exemplary directions for $\mathbf{k}$, $\mathbf{v_G}$, $\mathbf{H_{ext}}$ and the propagation angle $\theta= \angle(\mathbf{H_{ext}},\mathbf{v_G})= \arctan(dk_\mathrm{y}/dk_\mathrm{x})$ are shown in the graph. (b) Radiation direction $\theta$ calculated from the iso-frequency curves shown in (a). Dash-dotted lines correspond to the angles $\theta_{\mathrm{exp}}$ observed in the experiment.}
	\label{iso_graphen}
\end{figure}

For given frequency and external field, the iso-frequency curve $f(k_\mathrm{x}, k_\mathrm{y})=const$ can be calculated analytically from the dispersion relation. Calculations for $f=2f_\mathrm{e}=7.0$\,GHz and $f=3f_\mathrm{e}=10.5$\,GHz are illustrated in Fig.\,\ref{iso_graphen}(a), where $k_\mathrm{y}$ is shown as a function of $k_\mathrm{x}$. Using this data, we calculate the direction $\theta$ of the flow of energy of the spin waves relative to the externally applied field by:

\begin{equation}
	\theta= \angle(\mathbf{H_{ext}},\mathbf{v_G})=\arctan(v_x/v_y)= \arctan(dk_\mathrm{y}/dk_\mathrm{x}).
\end{equation}
 
Figure\,\ref{iso_graphen}(b) shows the calculated propagation angle $\theta$ in the CMFS waveguide as a function of $k_\mathrm{y}$. The most important feature in the trend of $\theta$ is the small variation of $\Delta \theta \leq 2^{\circ}$ in the range of $k_y=2-7$\,\mum{}$^{-1}$ for both frequencies $f=2f_\mathrm{e}$ and $f=3f_\mathrm{e}$. While the wavevector $\mathbf{k}$ changes, the direction of $\mathbf{v_G}$ - and, thus, the flow of energy - keeps almost constant as a function of $\mathbf{k}$. In this range, which includes the maximum wavevector $k_\mathrm{max}\approx6.3$\,\mum{}$^{-1}$ that can be emitted from the edge mode (see considerations above), the calculations yield $\theta_{\mathrm{calc}}(2f)=79^{\circ}$ and $\theta_{\mathrm{calc}}(3f)=66^{\circ}$ as mean values, respectively. The dash-dotted lines in Fig.\,\ref{iso_graphen}(b) represent the propagation angles $\theta$ of the spin-wave beams observed experimentally as shown in Fig.\,\ref{intensity_maps} ($\theta_{\mathrm{exp}}(2f)=78^{\circ}$ and $\theta_{\mathrm{exp}}(3f)=67^{\circ}$). The comparison of experimental findings and analytical calculations shows an agreement within the expected accuracy of our measurement setup and is, therefore, supporting our conclusion. Higher harmonics with $k_\mathrm{y} \leq 2$\,\mum{}$^{-1}$ are emitted with strongly varying directions from the edge mode and can be regarded as a negligible background in our measurement.

In summary, we reported nonlinear higher harmonic generation from a localized source in a micro-structured CMFS waveguide leading to the emission of strongly directed spin-wave beams or caustics. This observation results from the complex interplay of different phenomena in magnonic transport in magnetic microstructures. The localization of an edge mode due to demagnetizing fields in the waveguide leads to the formation of a source for the following nonlinear processes. The nonlinear higher harmonic generation results in the resonant excitation and emission of propagating spin waves at $2f_\mathrm{e}$ and $3f_\mathrm{e}$ in a wavevector range corresponding to the localization of the edge mode. The experimentally observed power dependencies of the different spin-wave modes show the expected behavior for direct resonant excitation and nonlinear higher harmonic generation. As shown by our calculation, the anisotropic spin-wave dispersion yields a well-defined direction of the flow of energy of the emitted spin waves in the relevant range in k-space. The calculation is not only qualitatively in accordance with our experimental findings but does also show quantitative agreement.

We gratefully acknowledge financial support by the DFG Research Unit 1464 and the Strategic Japanese-German Joint Research from JST: ASPIMATT. Thomas Br\"{a}cher  is supported by a fellowship of the Graduate School Materials Science in  Mainz (MAINZ) through DFG-funding of the Excellence  Initiative (GSC 266).  We thank our colleagues from the \textit{Nano Structuring Center} of the TU Kaiserslautern for their assistance in sample preparation.

\end{document}